\documentclass[preprint,superscriptaddress,double-spaced,floatfix,longbibliography]{revtex4-1}
\usepackage{graphicx,amsmath,bbm,bm,array,subfigure}
\usepackage{appendix}
\usepackage{lipsum}
\usepackage{dsfont}
\usepackage{calrsfs}
\usepackage{lineno}   
\usepackage{bbold}
\usepackage{relsize}
\usepackage[linktocpage=true,colorlinks=true,linkcolor=blue,citecolor=blue,urlcolor=blue]{hyperref}
\usepackage{MnSymbol}

\def\vec#1{\mathchoice
{\mbox{\boldmath $#1$}}
{\mbox{\boldmath $#1$}}
{\mbox{\boldmath $\scriptstyle #1$}}
{\mbox{\boldmath $\scriptscriptstyle #1$}}
}
\def \beq{\begin{equation}}
\def \eeq{\end{equation}}
\def \beqa{\begin{eqnarray}}
\def \eeqa{\end{eqnarray}}
\def \pd{\partial}
\def \nn{\nonumber}
\begin{document}
\begin{abstract}
We investigate the propagation of nonlinear energy density waves in a nonextensive quark-gluon plasma under the influence of a magnetic field using the reductive perturbation technique. For a nonextensive MIT bag equation of state, we obtain the governing equation for the first order perturbation of the energy density. We observe that an increase in the strength of the magnetic field results in the localization of the waves. 
\end{abstract}
\title{Nonlinear Pulse Dynamics in a Magnetized and Nonextensive Quark-Gluon Plasma}
\author{Trambak Bhattacharyya}
\email{trambak.bhattacharyya@gmail.com }
\affiliation{Bogoliubov Laboratory of Theoretical Physics, Joint Institute for Nuclear Research, Dubna 141980, Moscow Region, Russian Federation} 
\author{Md Hasanujjaman}
\email{jaman.mdh@gmail.com}
\affiliation{Department of Physics, Darjeeling Government College, Darjeeling- 734101, India}
\maketitle
\section{Introduction}
\label{sec1}
Studying the properties of the hot and dense Quark-Gluon Plasma (QGP), created in high-energy collisions at the LHC at CERN, or the RHIC at BNL, is an active field of research. Some of the important research directions in the field of high-energy collisions involve different types of evolution that take place after collisions. On one hand, the system of quarks and gluons created after a collision is an evolving one. On the other hand, the distribution of the high-energy particles that are created before the formation of the medium also evolves, when it passes through the medium. The evolution of the hot and dense medium is studied using the relativistic hydrodynamic equation, and the evolution of high-energy particles (also known as `jets') inside the QGP may be investigated with the help of the Boltzmann Transport Equation (BTE). Jets lose energy inside the medium and create energy density perturbations that travel as nonlinear waves through the medium. The evolution of such nonlinear waves has been the subject matter of study in a few research articles \cite{Raha1,Raha2,Fogaca:2009wf,Fogaca:2011pk,Fogaca:2014gwa,Bhattacharyya:2020sua,Sarwar:2020oux,Sarwar:2021csp}.

The basic mathematical equation governing the evolution of perturbations in an ideal fluid is given by Euler's equation. Additionally, one has to take into account the continuity equation of the entropy density. However, to solve this set of equations one needs to introduce the equations of state (EoS) that take into account the relationships among the macroscopic variables characterizing the medium. One possibility may be simply to consider the conventional Boltzmann-Gibbs statistics and calculate the thermodynamic variables like energy density, entropy density, and pressure.
For example, in Ref.~\cite{Fogaca:2009wf}, authors considered the MIT bag EoS of an ideal gas of massless quarks and gluons and obtained breaking wave solutions. The authors also comment that whether a medium will support soliton propagation (as opposed to yielding a breaking wave solution) or not will depend on the equation of state. Such an example of soliton propagation has been observed in a cold quark-gluon plasma using the QCD mean-field approach to derive the equation of state \cite{Fogaca:2011pk}.

However, the Boltzmann-Gibbs statistics may not represent the physical systems under consideration that have a fluctuating ambiance. Fluctuations in the positions of the nucleons inside the colliding nuclei may give rise to energy density and hence, temperature fluctuation. It has been shown that in a fluctuating ambiance power-law distributions (as opposed to the exponential Boltzmann-Gibbs distributions) appear, and power-law distributions originating from the Tsallis nonextensive statistics have been extensively used in many fields of research. Experimental signatures of such a scenario are found in the particle spectra that follow the nonextensive power-law statistical distributions \cite{CMSTs1,ALICETs1}. These distributions are characterized by the nonextensivity parameter $q$ that is related to the relative variance of inverse temperature \cite{Wilkprl}. 
The evolution of nonlinear perturbations in such a fluctuating ambiance has been the subject matter of our previous works \cite{Bhattacharyya:2020sua,Sarwar:2021csp} that considered the EoS$^{\text{s}}$ inspired by the nonextensive statistics.
In the present article, we study the evolution of nonlinear perturbation waves under the influence of a magnetic field. A transient magnetic field of about $eB\sim (1-10) m_{\pi}^2$ is generated in non-central high-energy collisions \cite{Bzdak:2011yy,PhysRevC.85.044907}, and it will be realistic to consider the magnetohydrodynamic equations to study the evolution of the perturbation waves. To solve the magnetohydrodynamic equation, one needs the EoS that can be calculated considering the nonextensive background under the influence of a large magnetic field. The present work considering relativistic magnetohydrodynamics is expected to generalize the previous works \cite{FogacamagNPA,FogacamagNLSCI} that focus on wave propagation in a non-relativistic cold quark-gluon plasma medium. 

During the analysis, we also computed analytical forms of thermodynamic variables of a nonextensive medium under the influence of a magnetic field to establish the equations of state. As far as our knowledge goes, such analytical forms have not been computed in any other works so far, and they may be important inputs to other related studies like examining the properties of nonextensive quantum gas exposed to a high magnetic field. One of the conclusions our analysis draws is that magnetic fields can help stabilize nonlinear energy density perturbation inside a hot QGP, in the same way it stabilizes baryon density perturbation in a cold QGP \cite{FogacamagNLSCI}.

The paper is organized as follows. The next section will be devoted to describing the mathematical model for the study. Section \ref{results} discusses the findings. Finally, we summarize, conclude, and provide an outlook of the results in Section \ref{summary}.

\section{Mathematical model}
\subsection{Magnetohydrodynamics and the equation for perturbation}
The evolution dynamics of the QGP (and any perturbation generated inside) are governed by the hydrodynamic equations. The order of the theory of hydrodynamics is dictated by the truncation of the order of the energy-momentum tensor (EMT) of the dissipative fluxes. Taking up to zeroth order of the dissipative fluxes are well described by the ideal hydrodynamics, and the governing equations are known as Euler's equation. Similarly, the first-order~\cite{Eckart:1940te,Landau_fluid_mechanics} and the second-order theory~\cite{Israel:1979wp,Muronga:2001zk} can be derived by truncating the EMT up to the first and the second order of the dissipative fluxes respectively. The first-order theory (Navier-Stokes) is acausal and numerically unstable to correctly describe the fluid. However, there are a few papers that argue that first-order theory can be causal and stable too~\cite{Bemfica:2019knx,Bemfica:2020zjp,Kovtun:2019hdm,Das:2020fnr}. Nevertheless, as the second-order theory is free from causality and stability problems, it is used as a conventional theory to describe the medium.  However, for the sake of simplicity, in this paper, we use the one-dimensional Euler's equation in the presence of a constant magnetic field to study the propagation of nonlinear waves in QGP.

 Throughout the article, we have considered  $\hbar=c=K_B=1$, and the choice of the metric is $g^{\mu \nu}$=(+,\,-,\,-,\,-), such that the fluid's four-velocity follows $u^\mu u_\mu=1$. The four-velocity is written as $u^\mu=\gamma(1,\vec{v})$, where $\gamma=(1-v^2)^{-1}$.

In an off-central collision, a huge transient magnetic field of order up to $10^{17}-10^{19}$ Gauss is produced. Although there will be an electric field due to this transient behavior, we do not consider it in the present study. The Euler's equation in the presence of a magnetic field is given by~\cite{Roy:2015kma}:
\begin{equation}
\label{Euler}
\frac{\pd \vec{v}}{\pd t}+(\vec{v}.\vec{\nabla})\,\vec{v}=-\frac{1}{\gamma^2(\epsilon+P+B^2)}\Big[\vec{\nabla} \big(P+\frac{B^2}{2}\big)+\vec{v}\,\frac{\pd}{\pd t}\big(P+\frac{B^2}{2}\big)\Big]\,,
\end{equation}
where, $\epsilon$ is the energy density, $P$ represents the pressure of the fluid, and $B$ is the strength of the magnetic field. 
For the sake of simplicity of the calculation, we have considered a uniform magnetic field in Euler's equation to study the propagation of the nonlinear waves. 

Along with the Euler's equation, the fluid follows the continuity equation of the entropy density $s$ given by:
\begin{equation}
\label{Continuity}
\frac{\pd {s}}{\pd t}+\gamma^2v s\,\Big[\frac{\pd \vec{v}}{\pd t}+(\vec{v}.\vec{\nabla})\,\vec{v}\Big]+\vec{\nabla}.(s\vec{v})=0.
\end {equation}
In the present article, we are interested in studying the perturbations in energy density. We also assume that the perturbation propagates along a preferred direction (beam direction) $x$ \cite{Fogaca:2009wf}. As the perturbation of the energy density may be comparable to the background energy density, the linearization technique may become insufficient. To take this possibility into account, the reductive perturbative technique (RPM) has been used in which the energy density and the velocity profile has been expanded in terms of an expansion parameter $\sigma$ as~\cite{Fogaca:2009wf,Sarwar:2021csp}:
\beqa
\epsilon&=&\epsilon_0(1+\sigma\,\epsilon_1+\sigma^2\,\epsilon_2+\sigma^3\,\epsilon_3+.\,.\,.)\,\,\,\,\,\,\,\,\,\,\rightarrow \,\,\,\,\,\hat{\epsilon}=\frac{\epsilon}{\epsilon_{0}}=1+\hat{\epsilon}_{1}+\hat{\epsilon}_{2}+\hat{\epsilon}_{3}+.\,.\,\nn\\
P&=&P_0(1+\sigma\,P_1+\sigma^2\,P_2+\sigma^3\,P_3+.\,.\,.)\,\,\,\rightarrow \,\,\,\hat{P}=\frac{P}{P_{0}}=1+\hat{P_{1}}+\hat{P}_{2}+\hat{P}_{3}+.\,.\,\nn\\
v&=&c_s(\sigma v_1+\sigma^2 v_2+\sigma^3 v_3+.\,.\,.)\,\,\,\,\,\,\,\,\,\,\,\,\,\,\,\,\,\,\,\,\,\rightarrow \,\,\,
\hat{v}=\frac{v}{c_{s}}=\hat{v_{1}}+\hat{v}_{2}+\hat{v}_{3}+.\,.\,.\,
\eeqa
In the above equation, $c_s$ is the speed of sound. We also neglect the $\mathcal{O}(\sigma^3)$ terms and get:
\beqa
({1+c_s^2})\left[\frac{\pd \epsilon}{\pd t}+v\left(\frac{\pd \epsilon}{\pd x} \right) \right]+({\epsilon+P+B^2})\left[\left(\frac{\pd v}{\pd x} \right)+v\left(\frac{\pd v}{\pd t} \right)+ v\left(\frac{\pd v}{\pd x} \right) \right]=0\,.
\label{eq4}
\eeqa
where we have used $c_s=(\pd P/\pd \epsilon)$. Now we 
use a coordinate transformation to the stretched coordinate system given by $(x,\,t)\rightarrow (\xi,\,\tau)$ as: 
\begin{equation}
\xi=\frac{\sigma^{1/2}}{L}(x-c_s t);\,\,\,\tau=\frac{\sigma^{3/2}}{L} c_s t\,,
\end{equation}
where $L$ is some characteristic length. By collecting different orders of $\sigma$ and putting them to zero, we obtain:
\beqa
v_1=\frac{c_s(1+c_s^2)\epsilon_0}{\epsilon_0+P_0+B^2}\,\epsilon_1
\eeqa

Now, using Eq.~\eqref{eq4} and reverting to $(x,\,t)$ coordinate, we get the governing equation for the first-order perturbation of the energy density as follows:
\beqa
\frac{\pd \hat{\epsilon}_{1}}{\pd t}+c_s\Bigg[1+\frac{(1+c_{s}^{2}) \epsilon_{0}}{\epsilon_{0}+P_{0}+B^{2}}\,\hat{\epsilon}_1+\frac{\epsilon_0 }{\epsilon_0+P_0+B^2}\hat{\epsilon}^2_1\Bigg\{\Big(1+\frac{P_{0}}{\epsilon_{0}}c_{s}^{2}\Big) +c_{s}^{2}(1+c_{s}^{2})\Bigg\}\Bigg]\frac{\pd \hat{\epsilon}_{1}}{\pd x}=0\,.
\label{eom}
\eeqa
In the above equation, we need inputs like the background energy density $\epsilon_0$, background pressure $P_0$, and the velocity of sound $c_s$ that we estimate from the MIT bag model in the next section.

\subsection{Equations of state}
We consider the gluons and quarks inside the medium to be represented by the following quantum distributions derivable from the nonextensive statistical mechanics \cite{TBParvanEPJA2}.

\begin{eqnarray}
n_f=\frac{1}{\left[1+(q-1)\frac{E_p-\mu}{T}\right]^{\frac{q}{q-1}}+1}
\nn\\
n_b=\frac{1}{\left[1+(q-1)\frac{E_p-\mu}{T}\right]^{\frac{q}{q-1}}-1}
\label{TsFDBE},
\end{eqnarray}
$E_{p}=\sqrt{p^2+m^2}$ is the single particle energy of a particle of 3-momentum $\vec{p}$ and mass $m$, $\mu$ is chemical potential, $q$ is the nonextensivity parameter, and $T$ is temperature.

In this work we consider the nonextensive MIT bag model \cite{TsMITBag} that gives us (for the case of zero chemical potential)
\begin{eqnarray}
\epsilon_{\mathrm{bag}} &=& \mathcal{B}+\epsilon_{\text{b}}+2\epsilon_{\text{f}}, \label{bagepsilon}\\
P_{\mathrm{bag}} &=& -\mathcal{B}+P_{\text{b}}+2P_{\text{f}}, \label{bagP}
\end{eqnarray}

Thermodynamic variables like energy density ($\epsilon$) and pressure ($P$) appearing in Eq.~\eqref{bagP} are given in terms of the distributions in Eq.~\eqref{TsFDBE} by

\begin{eqnarray}
\epsilon_{i}= g\int\frac{d^3p}{(2\pi)^3}~E_p~n_i;~\label{epsilon}
P_i=g\int\frac{d^3p}{(2\pi)^3}\frac{p^{2}}{3E_p}~n_i, \label{P}
\end{eqnarray}
where $g$ is the degeneracy factor and $i=f,b$ stands for the fermions (quarks) and bosons (gluons).

However, due to the presence of a magnetic field, the energy values become quantized into the so-called Landau levels and the three-momentum integration becomes modified. Assuming a uniform magnetic 3-vector field along the z-direction $\Vec{B}=B \hat{z}$, the single particle energy of the $j$-th ($j$ $\in$ $\mathbb Z^{\geq}$) Landau level is given by \cite{Bannurmagtherm},
\begin{equation}
    E_{pj} = \sqrt{m^2+p_z^2+2j|q_f eB|},
\end{equation}
and the momentum integration undergoes following modification \cite{Bannurmagtherm}
\begin{equation}
    \int\frac{d^3p}{(2\pi)^3} \rightarrow \frac{|q_f eB|}{2\pi} \sum_{j=0}^{\infty} \int\frac{dp_z}{2\pi} \left(2-\delta_{0j}\right).
\end{equation}
where $2-\delta_{0j}$ is the degeneracy of the $j$-th Landau level.

\subsubsection{Gluons}
Gluons do not carry an electric charge and are not influenced by a magnetic field. The analytical forms of the nonextensive energy density and pressure of the gluonic subsystem are given by \cite{Bhattacharyya:2020sua}
\begin{eqnarray}
 P_\text{b} &=& \frac{g T^4}{6 \pi ^2 (q-1)^3 q} \left[3 \psi
   ^{(0)}\left(\frac{3}{q}-2\right) + \psi
   ^{(0)}\left(\frac{1}{q}\right)- 3 \psi ^{(0)}\left(\frac{2}{q}-1\right) - \psi
   ^{(0)}\left(\frac{4}{q}-3\right)\right], 
   \label{Pboson} \nonumber\\
   \epsilon_\text{b} &=& \frac{g T^4}{2 \pi ^2 (q-1)^3 q} \left[ 3 \psi
   ^{(0)}\left(\frac{3}{q}-2\right) + \psi
   ^{(0)}\left(\frac{1}{q}\right)- 3 \psi ^{(0)}\left(\frac{2}{q}-1\right) - \psi
   ^{(0)}\left(\frac{4}{q}-3\right) \right],
   \label{epsilonboson}
\end{eqnarray}
where $\psi^{(0)}$ is the digamma function \cite{Erdelyi}

\subsubsection{Quarks}
We consider the QGP medium to be formed also of (slightly) massive ($\sim$ 10 MeV) light quarks. The analytical forms of the nonextensive energy density and pressure of the subsystem with slightly massive fermions under the influence of magnetic field can be calculated using the Mellin-Barnes contour integration representation of the distributions given by Eq.~\eqref{TsFDBE}. The process has been detailed in Refs. \cite{TsMBPRD,TsMBMDPI}. 
In this work, we consider the lowest Landau level (LLL) approximation that implies considering $j=0$.

As we notice in Refs.~\cite{TsMBPRD,TsMBMDPI},
results can be divided into two regions of $q$ values
given by $q\geq 1+T/m$ (upper region, $\mu=0$), and $q< 1+T/m$ (lower region$, \mu=0$), as given below:

\noindent {\bf Upper region:} The pressure in the upper region is given by,
\begin{eqnarray}
P_f^{\text{(up)}}=\mathlarger{\mathlarger{\sum}}_{s=1}^{s_0}
(-1)^{s+1}g m |q_f e B|
\left[
\frac{m  
   \Gamma \left(\frac{q s}{2 \delta q}-1\right)
   \left(\frac{T}{\delta q m}\right)^{\frac{q s}{\delta q}} \, _2F_1\left(\frac{q s}{2
   \delta q},\frac{q s}{2
   \delta q}-1;\frac{1}{2};\frac{T^2}{m^2
   {\delta q}^2}\right)}{8 \pi ^{3/2} \Gamma
   \left( \frac{q s}{2 \delta q}+\frac{1}{2}\right)}
   \right.\nn\\
   -
   \left.
   \frac{ T \Gamma
\left(\frac{q s}{2\delta q}-\frac{1}{2}\right)
   \left(\frac{T}{\delta q
   m}\right)
   ^{\frac{q s}{\delta q}} 
   \, _2F_1
   \left(\frac{q s}{2\delta q}-\frac{1}{2}, 
   \frac{q s}{2 \delta q}+\frac{1}{2};\frac{3}{2};\frac{T^2}{m^2 \delta q^2}\right)}{4 \pi ^{3/2}
   \delta q \Gamma 
   \left(\frac{q s}
   {2 \delta q}\right)}
   \right].
   \label{Pup}
\end{eqnarray}
In the above equation, $\, _2F_1$ is the Hypergeometric function \cite{Erdelyi}, and $\delta q \equiv q-1$.

\noindent {\bf Lower region:} The pressure in the lower region is given by the analytic continuation (AC) of the two Hypergeometric functions in Eq.~\eqref{Pup} \cite{Erdelyi},

\begin{eqnarray}
P_f^{\text{(low)}}=\mathlarger{\mathlarger{\sum}}_{s=1}^{s_0} 
(-1)^{s+1}
g m |q_f e B|
\left[
\frac{m  
   \Gamma \left(\frac{q s}{2 \delta q}-1\right)
   \left(\frac{T}{\delta q m}\right)^{\frac{q s}{\delta q}} \mathcal{H}^{(\text{1,AC})}}
   {8 \pi ^{3/2} \Gamma
   \left( \frac{q s}{2 \delta q}+\frac{1}{2}\right)}
   -
   \frac{ T \Gamma
\left(\frac{q s}{2\delta q}-\frac{1}{2}\right)
   \left(\frac{T}{\delta q
   m}\right)
   ^{\frac{q s}{\delta q}} 
   \mathcal{H}^{(\text{2,AC})}}{4 \pi ^{3/2}
   \delta q \Gamma 
   \left(\frac{q s}
   {2 \delta q}\right)}
   \right],
\nn\\
\end{eqnarray}
where
\begin{eqnarray}
\mathcal{H}^{(\text{1,AC})}
\equiv
&&\frac{\sqrt{\pi } \Gamma
   \left(\frac{3}{2}-\frac{q s}{\delta q}\right) \left(\frac{T^2}{{\delta
   q}^2 m^2}\right)^{-\frac{q s}{2
   \delta q}} \, _2F_1\left(\frac{q
   s}{2 \delta q}+\frac{1}{2},\frac{q
   s}{2 \delta q};\frac{q
   s}{\delta q}-\frac{1}{2};1-\frac{m^2 {\delta q}^2}{T^2}\right)}{\Gamma
   \left(\frac{1}{2}-\frac{q s}{2
   \delta q}\right) \Gamma
   \left(\frac{3}{2}-\frac{q s}{2
   \delta q}\right)}  
   \nn\\
&&+
   \frac{\sqrt{\pi } \Gamma \left(\frac{q
   s}{\delta q}-\frac{3}{2}\right)
   \left(1-\frac{T^2}{\delta q^2
   m^2}\right)^{\frac{3}{2}-\frac{q
   s}{\delta q}}
   \left(\frac{T^2}{\delta q^2
   m^2}\right)^{\frac{q s}{2 \text{$\delta
   $q}}-\frac{1}{2}} \,
   _2F_1\left(\frac{1}{2}-\frac{q s}{2
   \delta q},1-\frac{q s}{2
   \delta q};\frac{5}{2}-\frac{q
   s}{\delta q};1-\frac{m^2
   \delta q^2}{T^2}\right)}{\Gamma
   \left(\frac{q s}{2 \text{$\delta
   $q}}-1\right) \Gamma \left(\frac{q s}{2
   \delta q}\right)};
\nn\\
\mathcal{H}^{(\text{2,AC})}
\equiv
&&\frac{\sqrt{\pi } \Gamma
   \left(\frac{3}{2}-\frac{q s}
   {\delta q}\right) \left(\frac{T}{\delta q m}\right)^{1-\frac{q s}{\delta q}} \,_2F_1\left(\frac{q s}{2 \delta q}-1, 
   \frac{q s}{2 \delta q}-\frac{1}{2};\frac{q s}{\delta q}-\frac{1}{2};1-\frac{m^2 \delta q^2}{T^2}\right)}{2 \Gamma \left(1-\frac{q s}{2 \delta q}\right) \Gamma \left(2-\frac{q s}
   {2\delta q}\right)}
   \nn\\
   &&+
   \frac{\sqrt{\pi } \delta q^4 m^4
   \Gamma \left(\frac{q s}{\delta q}-\frac{3}{2}\right)
   \left(1-\frac{T^2}{\delta q^2
   m^2}\right)^{\frac{3}{2}-\frac{q
   s}{\delta q}}
   \left(\frac{T}{\delta q m}\right)^{\frac{q s}{ \delta q}} \, _2F_1\left(\frac{3}{2}-\frac{q
   s}{2 \delta q},2-\frac{q s}{2
   \delta q};\frac{5}{2}-\frac{q
   s}{\delta q};1-\frac{m^2
   \delta q^2}{T^2}\right)}{2 T^4
   \Gamma \left(\frac{q
   s}{2\delta q}-\frac{1}{2}\right)
   \Gamma \left(\frac{q
   s}{2\delta q}+\frac{1}{2}\right)}.
\nn\\
\end{eqnarray}
$s_0$ is the number of terms required for convergence of the summation.

So, the total fermionic pressure is given in terms of the Heaviside $\theta$-function by,
\begin{eqnarray}
P_f = P_f^{\text{(up)}} \theta \left(q-1-\frac{T}{m}\right) + 
P_f^{\text{(low)}} \theta \left(1+\frac{T}{m}-q\right).
\end{eqnarray}
Depending on the values of the parameters, $q$ can be greater or less than $1+T/m$, leading to a non-zero contribution from the upper or the lower region.
To calculate the equation of state, we also need to compute the energy density that can be obtained from the following relation (when $\mu=0$):
\begin{eqnarray}
    \epsilon_f = T \left(\frac{\partial P}{\partial T} \right)
    -P
\end{eqnarray}

Using these expressions of pressure and energy density, we can estimate $\epsilon_{\text{bag}}$, and $P_{\text{bag}}$ that we equate to $\epsilon_0$, and $P_0$ respectively. Also, the velocity of sound is calculated from the following expression
\begin{eqnarray}
    c_s^2 = \left(\frac{\partial P}{\partial \epsilon} \right).
\end{eqnarray}

\section{Results and discussion}
\label{results}
We aim to observe the propagation of a perturbation in energy density dictated by Eq.~\eqref{eom}. Thermodynamic variables like energy density and pressure that appear in the equation are evaluated from the non-extensive statistics. To solve such an equation, we have considered an initial profile of $\hat{\epsilon}_1$ as:
\beqa
\hat{\epsilon}_1=A\,\Big[\text{sech}\Big(\frac{x-x_0}{B}\Big)\Big]^2
\eeqa
\begin{figure}[h]
\centering
\includegraphics[width=7.5cm]{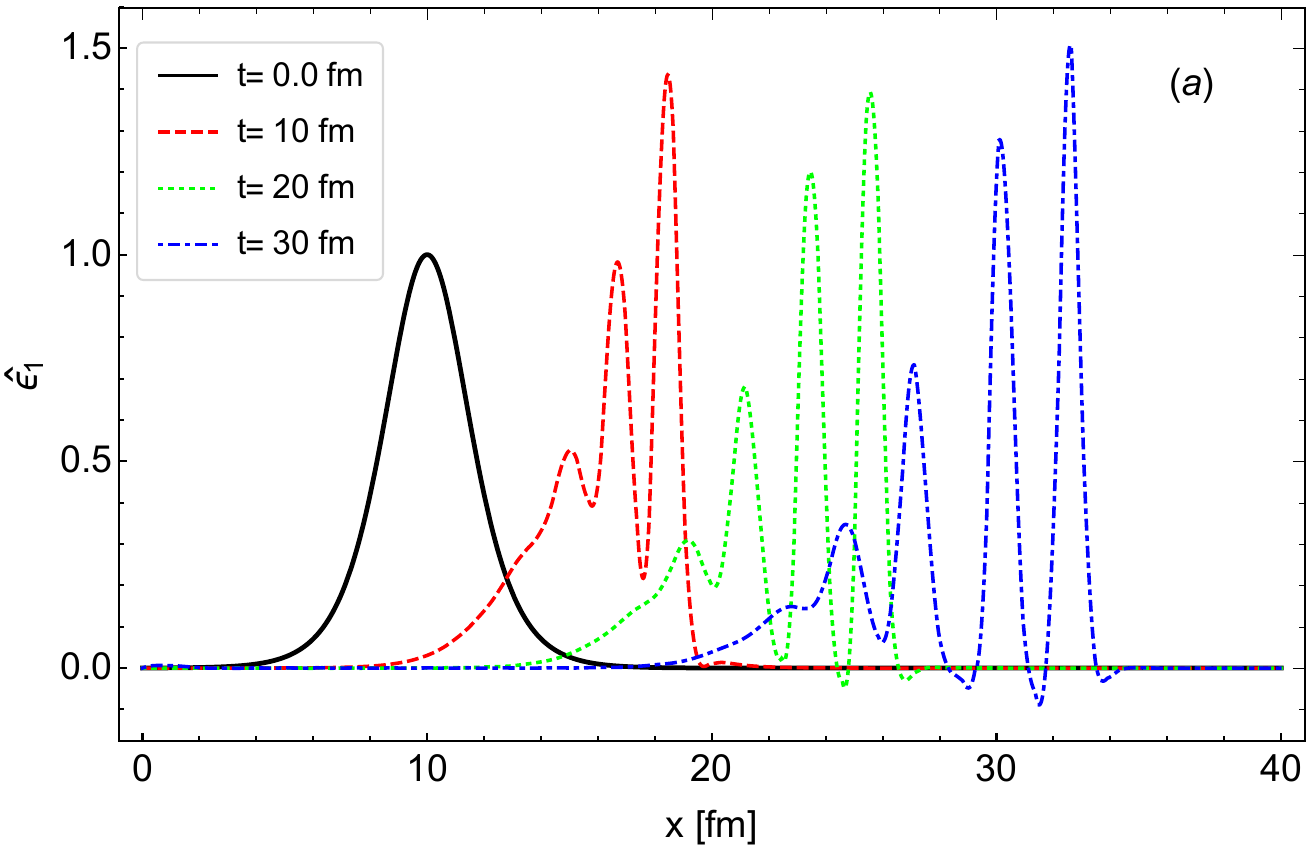}	
\includegraphics[width=7.5cm]{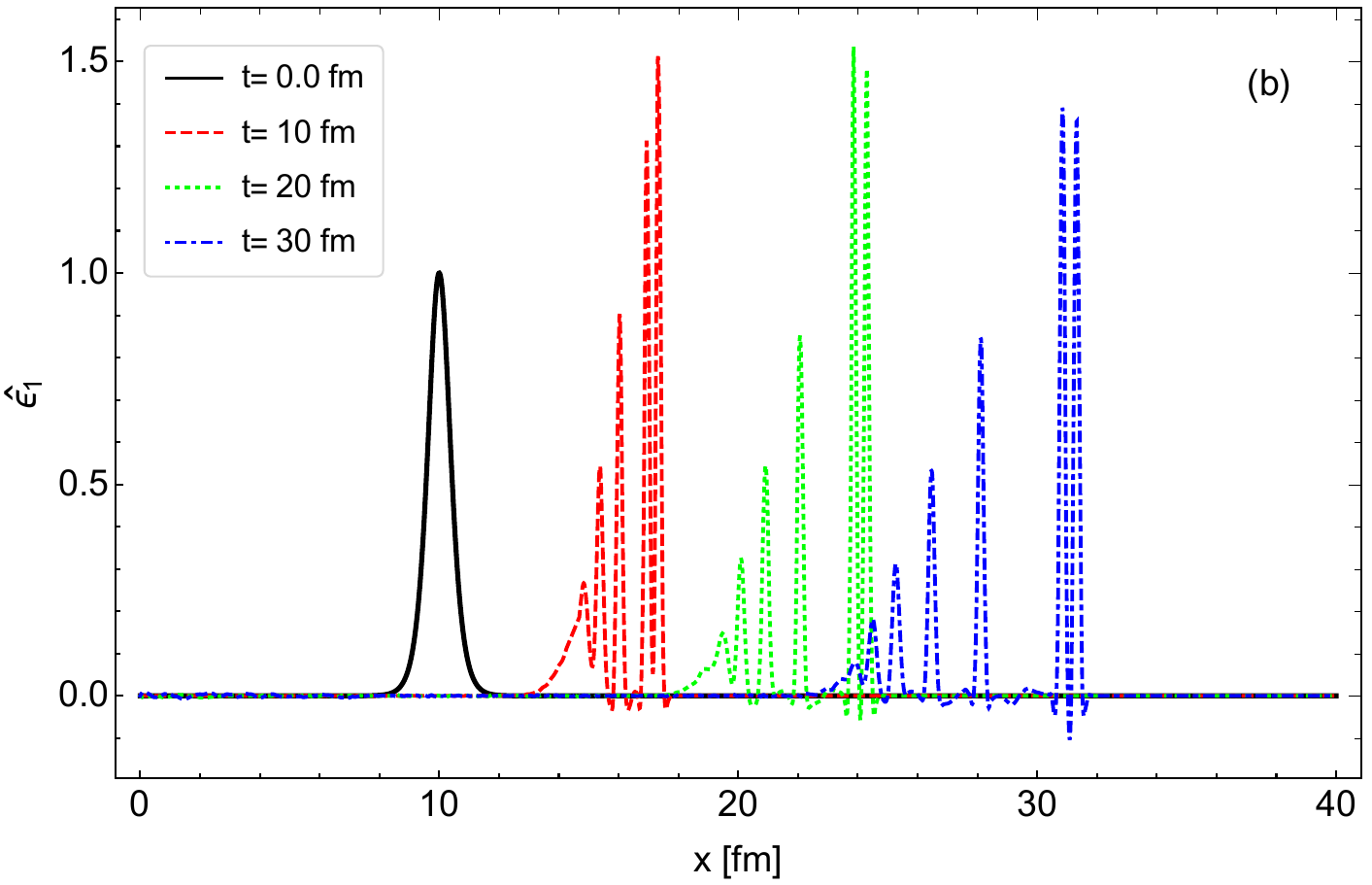}\\
\includegraphics[width=9.2cm]{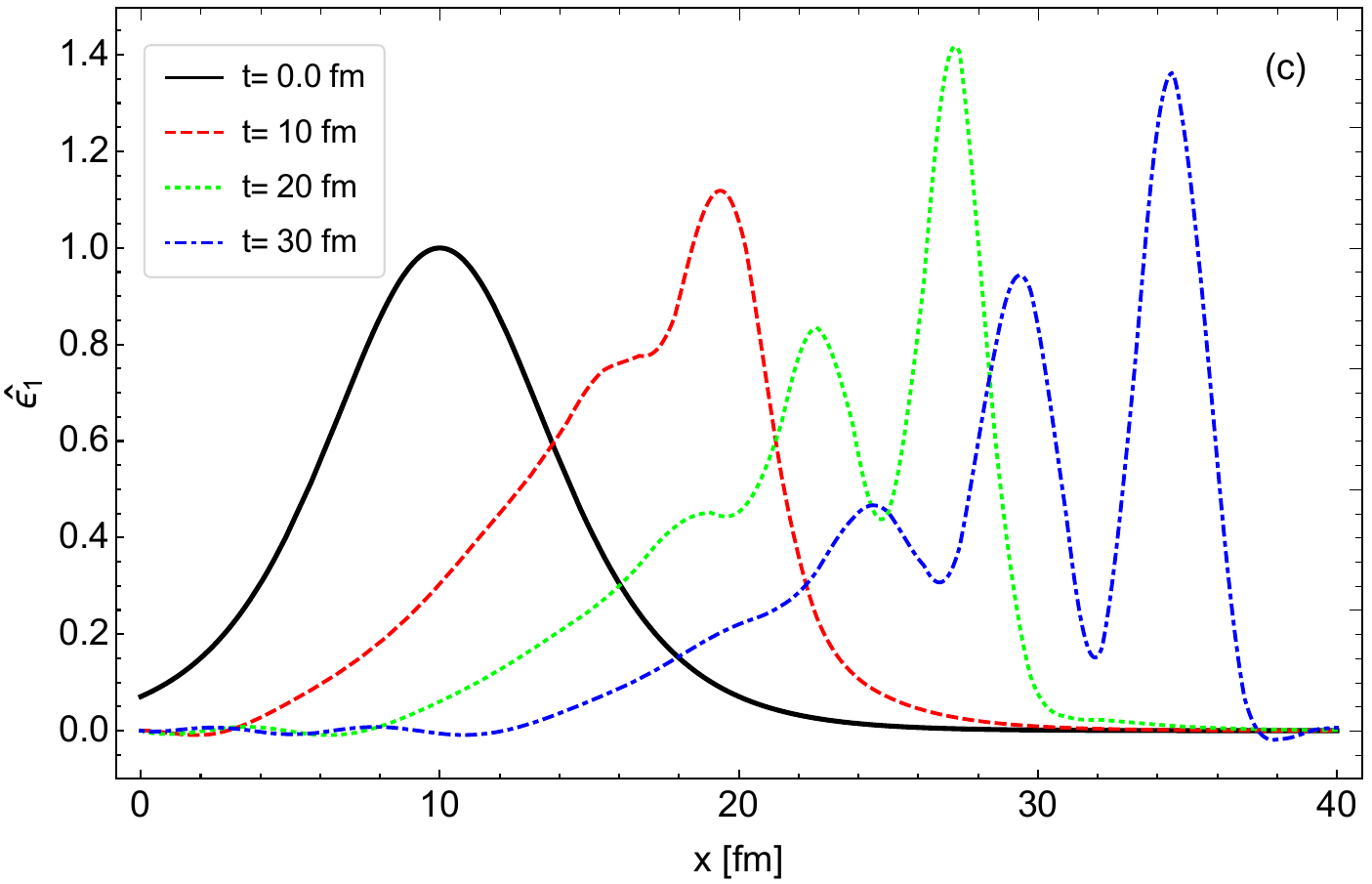}	
\caption{(Color online) (a) $\hat{\epsilon}_1$ as a function of $x$ at different time for 
$A=1$ and $B=2$ fm. 
(b) $\hat{\epsilon}_1$ as a function of $x$ at different time for $A=1$ and $B=0.5$ fm. 
(c) $\hat{\epsilon}_1$ as a function of $x$ at different time for $A=1$ and $B=5$ fm. The background temperature, $T=200$ MeV, and $q=1.10$ here. The strength of the magnetic field, $eB=m_\pi^2$.}
\label{fig1}
\end{figure}
\begin{figure}[h]
\centering
\includegraphics[width=9.5cm]{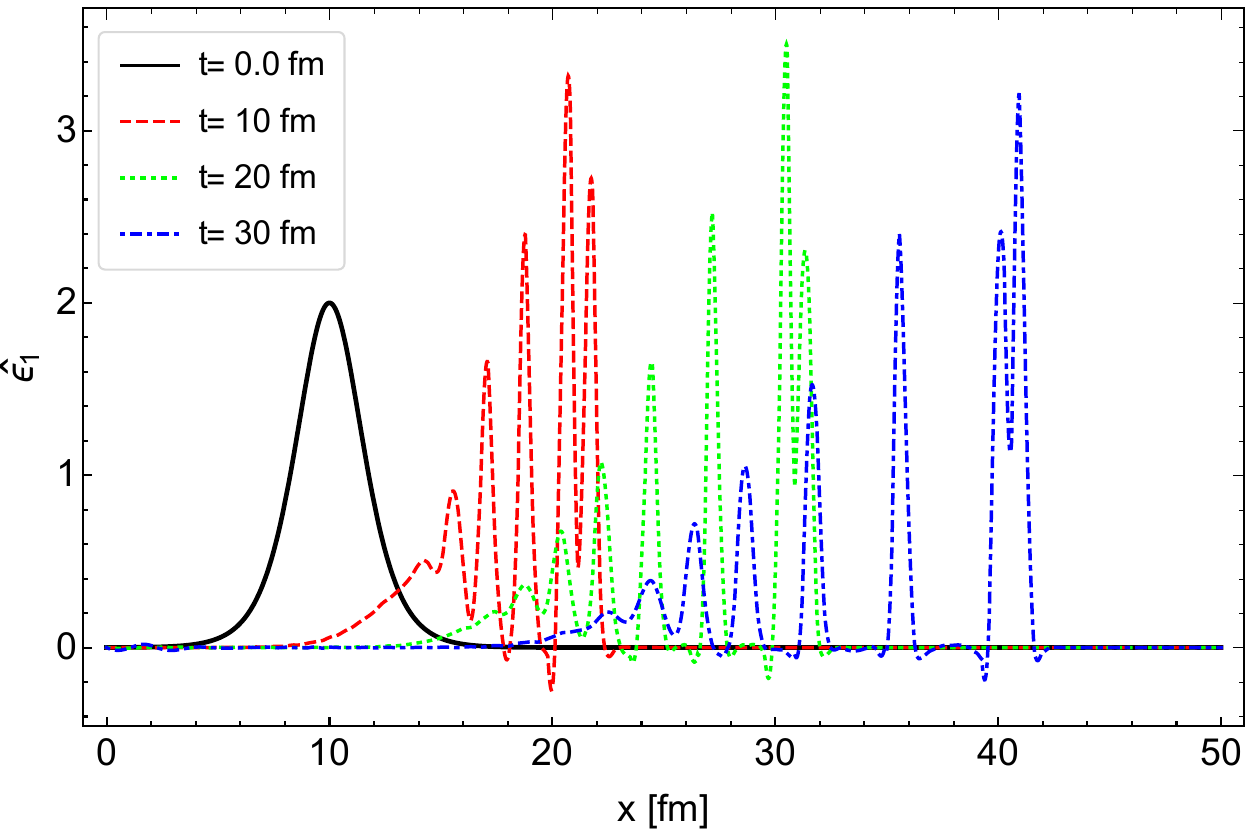}	
\caption{(Color online) $\hat{\epsilon}_1$ as a function of $x$ at different time for 
$A=2$ and $B=2$ fm. The background temperature,
$T=200$ MeV, the non-extensive parameter $q=1.10$, and the magnetic field strength $eB=m_\pi^2$.}
\label{fig2}
\end{figure}

where $A$ and $B$ respectively represent the magnitude and width of the profile. The $x_0$ is the peak of the initial position. Here we have taken $x_0=5$. The other choices of $x_0$ would not alter the desired result. 
Fig.~\ref{fig1} shows the propagation of $\hat{\epsilon}_1$ in QGP with a non-extensive ideal background of temperature $T=200$ MeV, the non-extensive parameter $q=1.10$, and the magnetic field strength $eB=m_\pi^2$.  Fig.~\ref{fig1}(a) is plotted with $A=1$ and $B=2$ fm at various times, where we see that with time the nonlinear pulses lose their localization with time. Fig.~\ref{fig1}(b) and Fig.~\ref{fig1}(c) are respectively plotted for ($A=1,\, B=0.5$ fm), and ($A=1,\, B=5$ fm) to see if the trend of the propagation remains same for different widths of the initial profile. Here we see a similar trend for the propagation of the waves, but the breaking of the waves is less for broader waves which is shown in Fig.~\ref{fig1}(c). To understand the propagation of the nonlinear wave with higher magnitude, we have considered $A=2$ keeping the same width $B=2$ fm. We observe that with the higher magnitude, the waves lose their localization early, which is shown in Fig.~\ref{fig2}.

\begin{figure}[h]
\centering
\includegraphics[width=9.5cm]{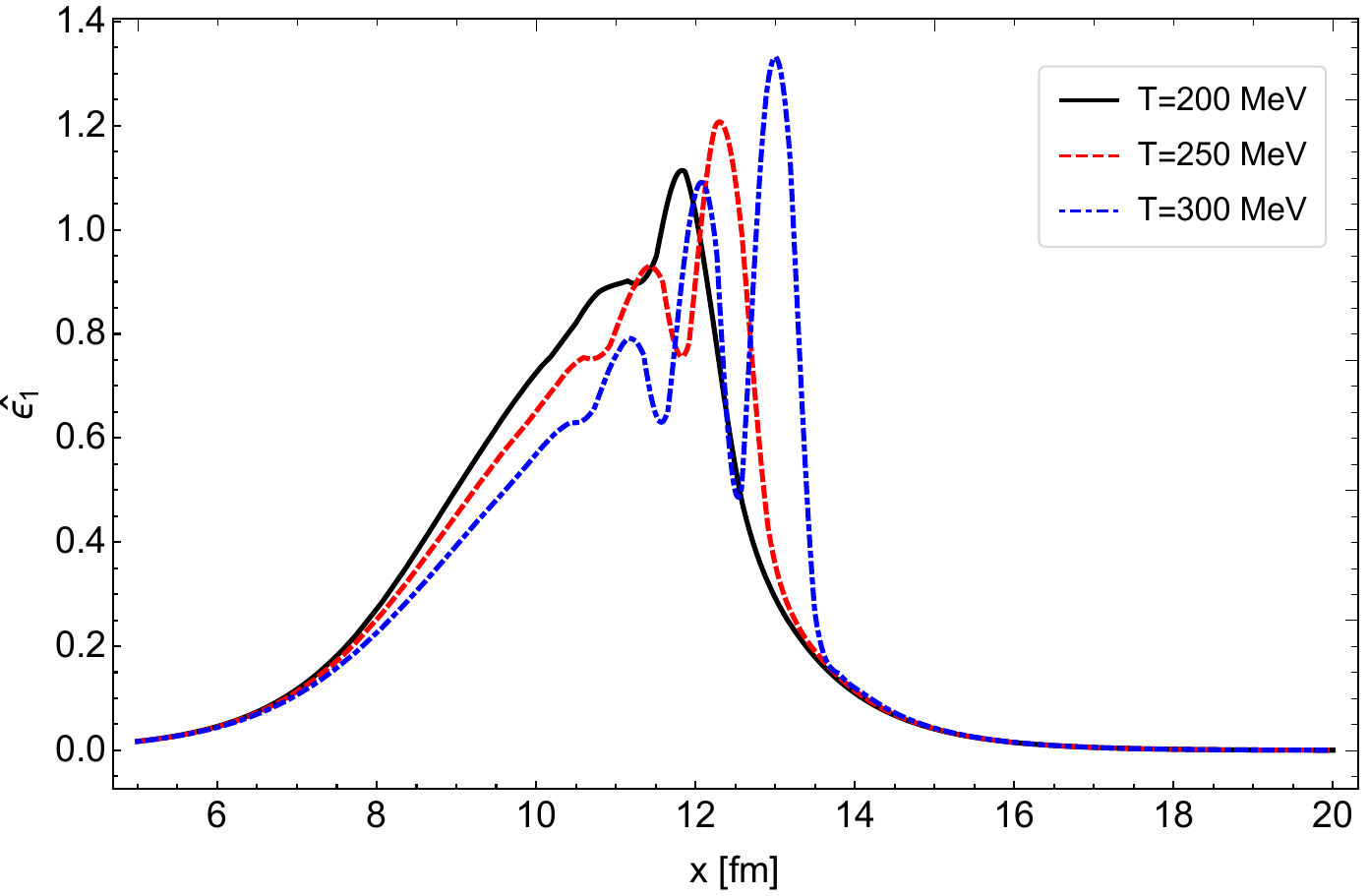}	
\caption{(Color online) $\hat{\epsilon}_1$ as a function of $x$ at different temperature when $t=1$ fm/c for 
$A=1$ and $B=2$ fm. The non-extensive parameter $q=1.10$, and the magnetic field strength $eB=m_\pi^2$.}
\label{fig3}
\end{figure}
\begin{figure}[h]
\centering
\includegraphics[width=9.5cm]{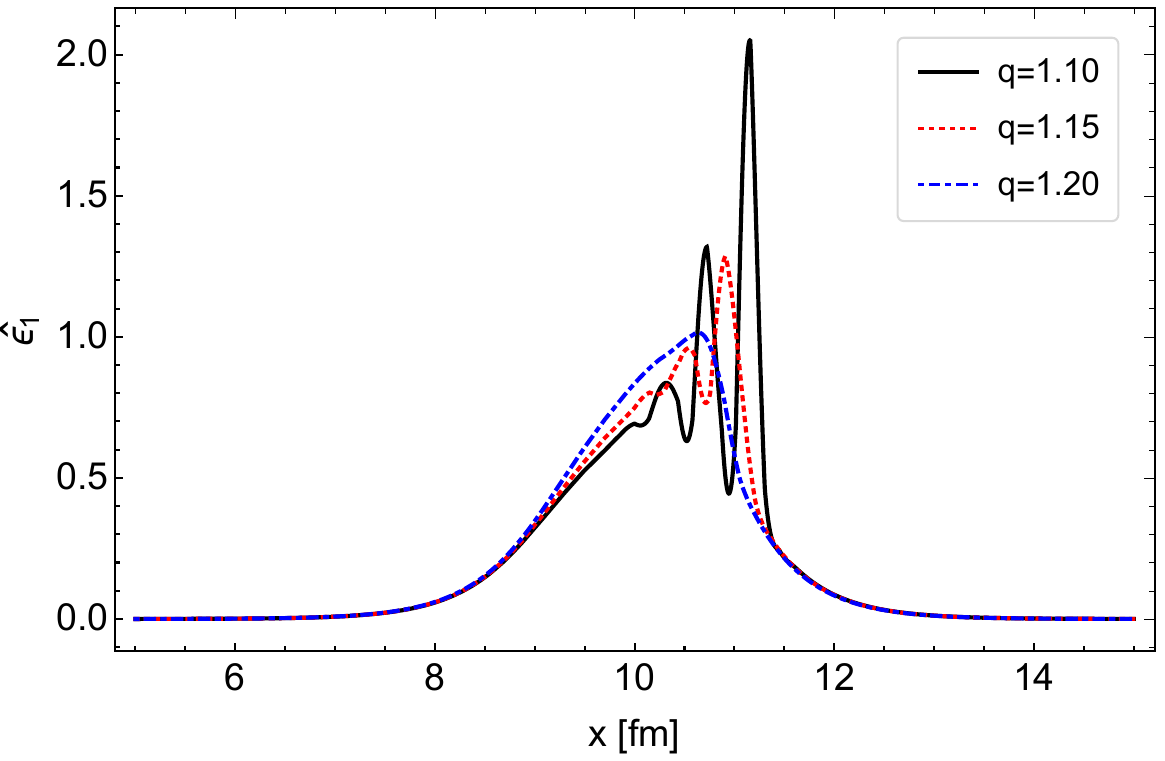}	
\caption{(Color online) $\hat{\epsilon}_1$ as a function of $x$ at different $q$ parameter values at $t=1$ fm/c for $A=1$ and $B=2$ fm. The background temperature, $T=200$ MeV, and the magnetic field strength $eB=m_\pi^2$.}
\label{fig4}
\end{figure}
In Fig.~\ref{fig3}, the effects of the background temperature on the propagation of the nonlinear waves are shown by considering the waves at $t=1$ fm/c for $q=1.10$, and the magnetic field strength $eB=m_\pi^2$. We see that with higher temperatures, the pulses are prone to break or delocalize in their position. Whereas, the non-extensive parameter effects are shown in Fig.~\ref{fig4}. Here, we have plotted $\hat{\epsilon}_1$ at $t=1$ fm/c for different values of $q$ parameter at $T=200$ MeV, when the strength of the magnetic field is $eB=m_\pi^2$. We observe that with an increase in the $q$ values, the waves tend to stabilize. 
\begin{figure}[h]
\centering
\includegraphics[width=7.5cm]{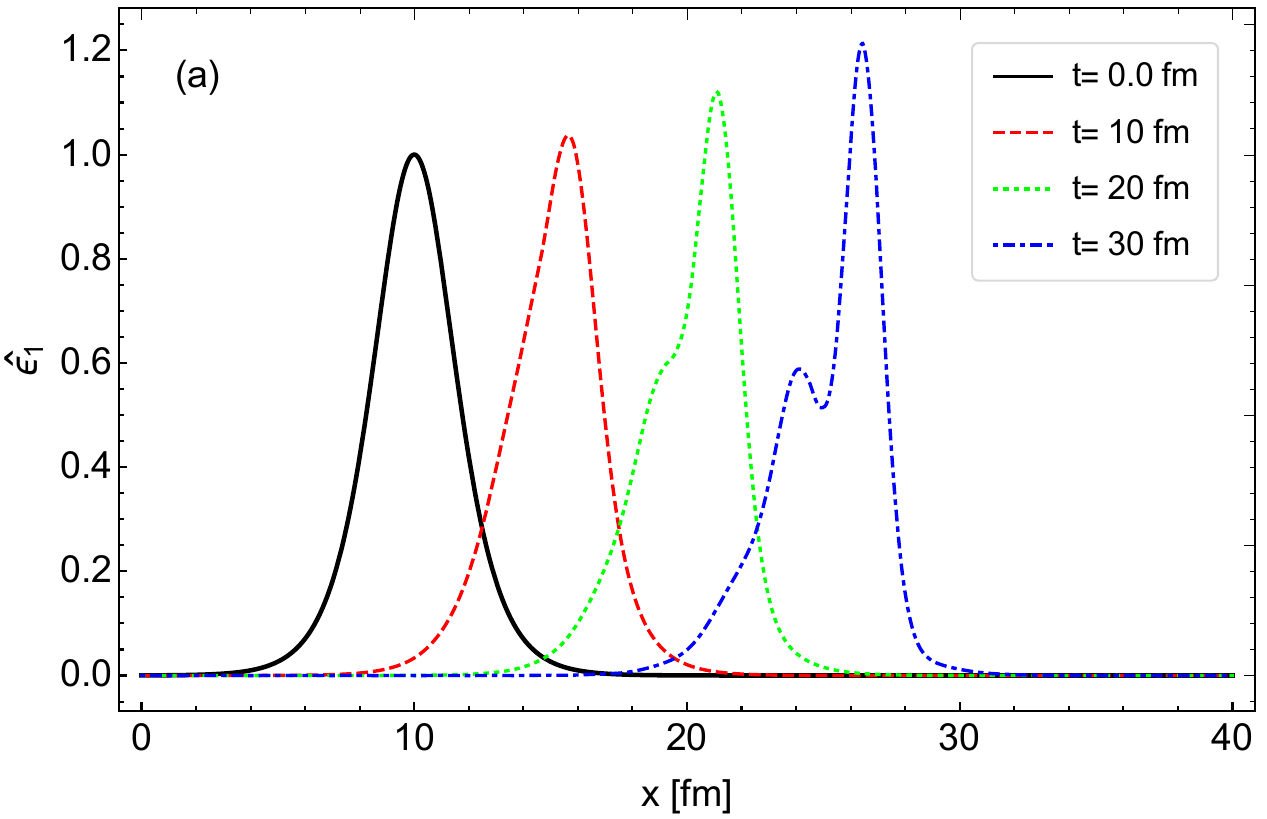}	
\includegraphics[width=7.5cm]{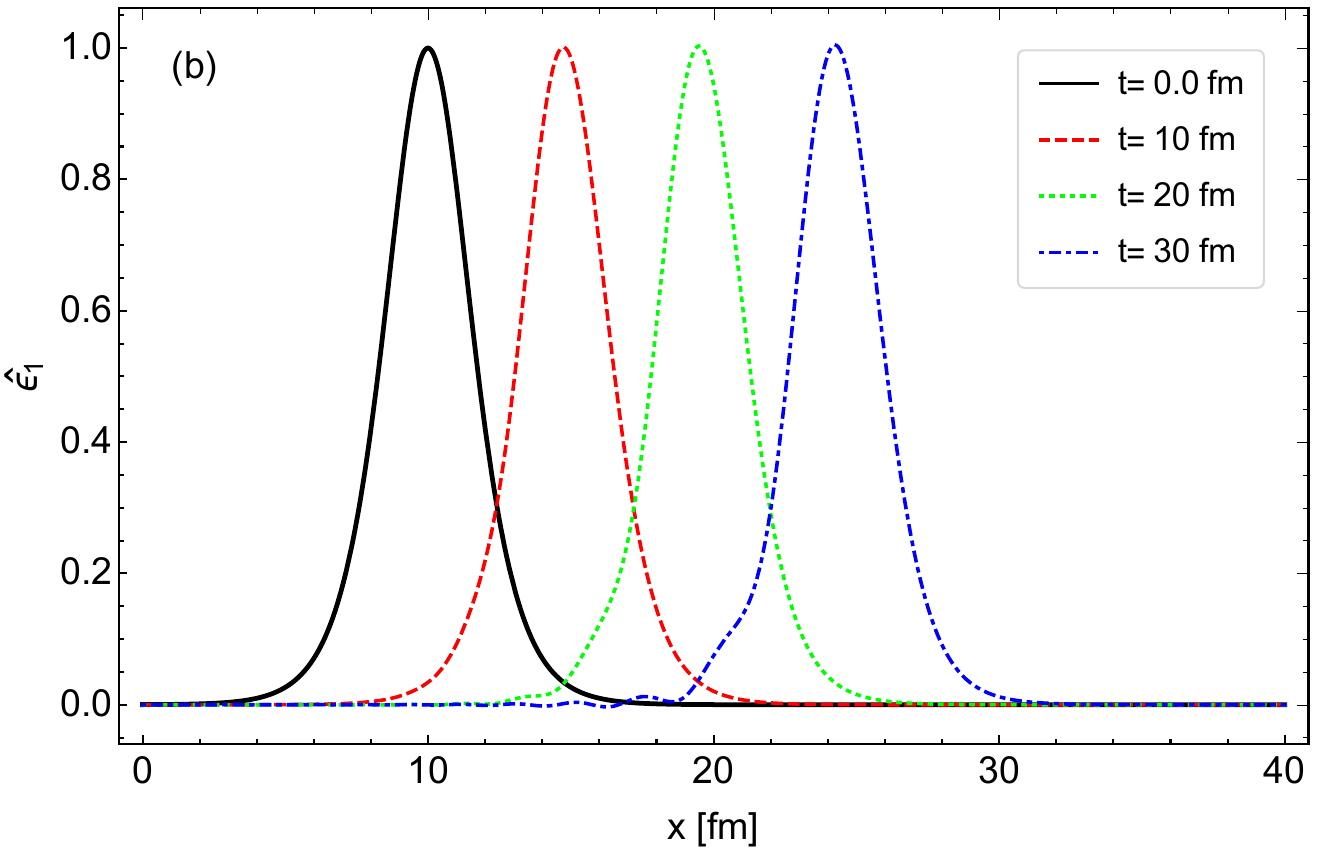}	
\caption{(Color online) $\hat{\epsilon}_1$ as a function of $x$ at different time for 
$A=1$ and $B=2$ fm. The background temperature,
$T=200$ MeV, the non-extensive parameter $q=1.10$. (a) the magnetic field strength $eB=4.4\,m_\pi^2$, and (b) The magnetic field strength $eB=11.5\,m_\pi^2$.}
\label{fig5}
\end{figure}
\begin{figure}[h]
\centering
\includegraphics[width=9.5cm]{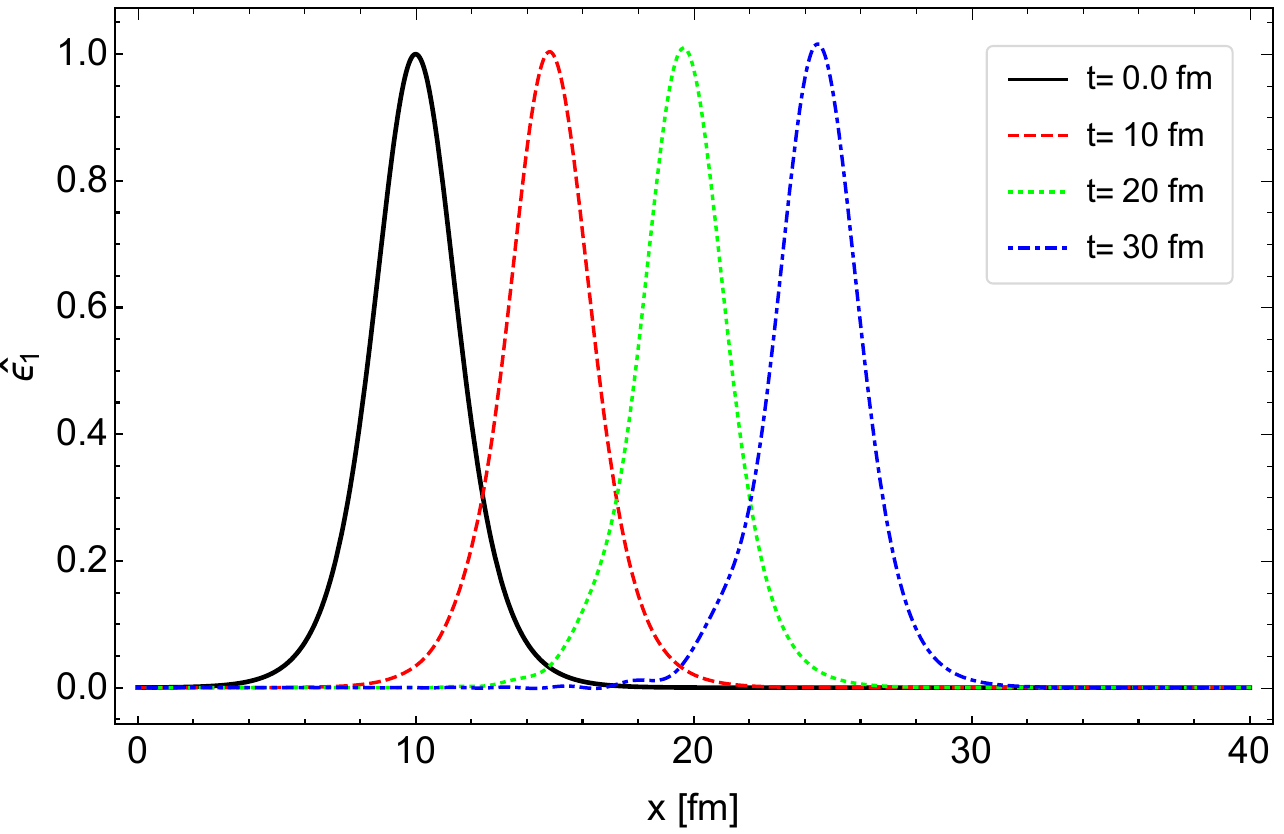}	
\caption{(Color online) (a) $\hat{\epsilon}_1$ as a function of $x$ at different time for 
$A=5$ and $B=2$ fm. The background temperature,
$T=200$ MeV, the non-extensive parameter $q=1.20$, and the magnetic field strength $eB=8.3\,m_\pi^2$.}
\label{fig6}
\end{figure}

The effects of the magnetic field are studied in Fig.~\ref{fig5}. The left panel is plotted for $T=200$ MeV and $q=1.10$ and with $eB=4.4\,m_\pi^2$. The right panel is for the same values of $T$ and $q$, but with $eB=11.5\,m_\pi^2$. We see that with the strengthened magnetic field, the waves regain their localization, which implies that the waves will survive within the lifetime of the QGP. Also, we notice that the shift of the peaks is less when the strength of the magnetic field is increased.

Another feature of the magnetic field is observed in the view of the $q$ parameter and is shown in Fig.~\ref{fig6}, where $\hat{\epsilon}_1$ is plotted for $T=200$ MeV, and $q=1.20$. It is shown that the waves just regain their stability at $eB=8.3\,m_\pi^2$, whereas in comparison with the right panel of Fig.~\ref{fig5}, the waves are just stabilized at $eB=11.5\,m_\pi^2$. 

\section{Summary, conclusions and outlook}
\label{summary}
In summary, we have considered the propagation of nonlinear waves inside a hot quark-gluon plasma under the influence of a uniform magnetic field. In this work, we have also utilized an equation of state inspired by the nonextensive statistical mechanics that arises for finite systems with fluctuations (in temperature, for example). Calculations of the equation of state have been performed assuming the lowest Landau level approximation. We find that with increasing amplitude and decreasing width of the initial perturbation lead to a greater degree of delocalization of the propagated wave. We further observe that with increasing $q$ parameter waves become more localized. However, a larger value of temperature results in more delocalization. At the LHC energy region, these two effects may compete with each other. It is further observed that with the increasing magnetic field, the waves become more localized. This behaviour is also reported in Ref.~\cite{FogacamagNLSCI} for baryon density perturbation in a cold quark-gluon plasma. This may be a consequence of the fact that with an increasing magnetic field, charged particles follow a helical path with a smaller radius, creating a perturbation with a smaller initial amplitude that develops into a more localized wave.

The present work may be benefitted from several generalizations that we reserve for the future. First of all, the present work may be generalized for a viscous quark-gluon plasma medium. Also, we may consider a changing magnetic field, keeping the magnetic flux fixed. This scenario will lead to anisotropic pressure \cite{GunnarJHEP}. It will also be interesting to study the effects of higher Landau levels in the equation of state. One should also consider the higher dimensional evolution of nonlinear waves to consider a more realistic physical scenario. Last but not the least, in this work we also obtained closed-form expressions for thermodynamic quantities in a gas dominated by fluctuations and exposed to a uniform magnetic field. These results may be important in studies related to quantum nonextensive gas in a high magnetic field.

\bibliography{NLW}
\end{document}